\documentclass[epj]{svjour}
\usepackage{graphics}
\usepackage{graphicx}
\usepackage{epsfig}
\usepackage{ulem}
\usepackage{morefloats}
\usepackage{rotating}
\usepackage{color,xcolor}
\usepackage{soul}
\usepackage{amsmath}
\usepackage{amssymb}

\def\beq{\begin{equation}}
\def\eeq{\end{equation}}
\def\beqa{\begin{eqnarray}}
\def\eeqa{\end{eqnarray}}
\def\ban{\begin{eqnarray*}}
\def\ean{\end{eqnarray*}}
\def\bi{\begin{itemize}}
\def\ei{\end{itemize}}

\definecolor{mypink}{RGB}{219, 48, 222}
\definecolor{brown}{RGB}{200, 150, 100}
\definecolor{green2}{rgb}{0.0, 0.5, 0.0}

\begin{document}
\title{Light clusters in warm stellar matter: calibrating the cluster couplings}
\author{Tiago Cust\'odio\inst{1} \and Alexandre Falc\~ao\inst{1} \and Helena Pais\inst{1} \and Constan\c ca Provid{\^e}ncia\inst{1} \and Francesca Gulminelli\inst{2} \and Gerd R\"opke\inst{3,4}
}                     
%
\mail{Helena Pais, \texttt{hpais@uc.pt}}
\institute{CFisUC, Department of Physics, University of Coimbra,
   3004-516 Coimbra, Portugal \and Normandie Univ., ENSICAEN, UNICAEN, CNRS/IN2P3, LPC Caen, F-14000 Caen, France \and Institut f\"ur Physik, Universit\"at Rostock, D-18051 Rostock, Germany \and National Research Nuclear University (MEPhI), 115409 Moscow, Russia}
\date{Received: date / Revised version: date}
%
\abstract{
The abundances of light clusters within a formalism that considers
in-medium effects are calculated using several relativistic mean-field
models, with both density-dependent and density-independent couplings. 
Clusters are introduced as new quasiparticles, with a modified 
coupling to the scalar meson field. A
comparison with experimental data from heavy ion collisions  allows
settling the model dependence of the results and the
determination of the 
couplings of the light clusters to the meson fields.   
 We find that extra experimental constraints at higher density are needed 
to convincingly pin down the density associated to the melting of clusters in the dense nuclear medium.
The role of neutron rich clusters, such as $^6$He, in asymmetric matter is discussed.
%
} 
\maketitle
\section{Introduction}
\label{intro}

Below saturation density, nuclear matter is supposed to undergo a
liquid-gas phase transition \cite{barranco80,muller95,borderie19}. Since in physical systems nuclear matter is electrically charged, the phase separation will produce clusterized matter. This behavior is directly reflected in several astrophysical sites, like core-collapse supernovae \cite{arcones08,ropke08,fischer13,furusawa13,furusawa17}, neutron star (NS) mergers \cite{bauswein13,fernandez13,just14}, and the inner crust of neutron stars \cite{ravenhall83,schneider13,horowitz14}. The form of the clusterized matter depends on temperature and isospin asymmetry.  In cold catalysed beta-equilibrium matter, as the one occuring in neutron stars, spherical clusters are found in almost the whole inner crust region, and close to the crust-core transition, the competition between surface and Coulomb forces gives rise to cluster configurations of different geometries  coined ``pasta phases'' \cite{ravenhall83}. These types of clusters may survive even at finite temperature \cite{Sonoda2007,Avancini2010,Avancini2017,Ji2020,Wu2017,Sedrakian2020}. Light clusters, i.e. light nuclei like deuterons or $\alpha-$particles, will form  in warm stellar matter as found in core-collapse supernova matter, proto-neutron stars or binary neutron star mergers, and  may also coexist with heavy clusters at densities above 10$^{-2}$ fm$^{-3}$, if the temperature does not overcome a few MeV. The presence of   light clusters  affects the rates of  the  reactions involving the weak force,
and therefore, may impact the supernova dynamics \cite{arcones08,Fischer20}. In the NS merger evolution, the $\alpha$ particles play an important role on the dissolution of the remnant torus of accreted matter that surrounds the central high-mass NS formed after the merging. Matter from this accretion disk is also contributing to the  ejecta that originates the kilonova observation \cite{fernandez13,just14,rosswog15}.

Clusterized warm  matter  at low densities has been described within a
generalization of the relativistic mean-field (RMF) approach. Within this framework, light
clusters  are included as
independent degrees of freedom that interact  with the medium through
their coupling to the mesonic fields
 \cite{avancini10,avancini12,typel10,ferreira12,pais15,avancini17}.
In previous papers \cite{PaisPRC97,PaisPRC2019}, some of the present authors introduced a new formalism,
that takes into account in-medium effects for the calculation of the
equation of state with light clusters for  applications in
astrophysical systems.
These effects are introduced via the scalar cluster-meson coupling, and
also via an extra term that is added to the total binding energy of
the clusters. This term not only avoids double counting of
  single particle continuum states,  but also affects the dissolution of clusters at high densities. In both
references, the studies have been developed within the FSU model
\cite{FSU}, which has been fitted to both static and dynamic
properties, and it is adequate to describe nuclear matter at saturation
density and below. This model, however, is too soft and does not
describe neutron  stars with two solar masses. It is, therefore,
necessary to generalize the previous studies to other models
currently used with success to describe symmetric and asymmetric
nuclear matter.

 Some RMF models are frequently used in
simulations and in the study of astrophysical observations, such as the  RMF models with
non-linear mesonic terms  TM1 \cite{tm1,shen}
and its modifications \cite{providencia13,bao14,pais16,shen2020}, NL3 \cite{nl3} 
and its modifications \cite{horowitz01,pais16}, 
SFHo \cite{sfho}, FSU2R  \cite{FSU2R}, and RMF with density dependent
couplings, like DD2 \cite{typel10} and DDME2 \cite{ddme2}. Models such as
TM1 and NL3 have been fitted to the ground state properties of nuclei. However, they both present a too large slope of the symmetry energy
at saturation,  far from what  the experimental constraints
\cite{tsang12,lim13,oertel18}, or ab-initio chiral effective field
theory calculations (CEFT) \cite{hebeler13},   indicate, and, therefore, they
have been modified by a non-linear $\omega\rho$ term that smoothens the density dependence of the symmetry energy
\cite{horowitz01,providencia13,bao14,pais16,shen2020}. The
parametrization  FSU2R  \cite{FSU2R} is based in the FSU
model, which was modified in order to be able to describe two
solar mass stars and  still to satisfy CEFT results.

In \cite{PaisPRC97,PaisPRC2019}, it was found that the equilibrium constants
determined from the NIMROD data \cite{qin12}
could be well described taking a universal coupling  of the
$\sigma$-meson to all the light
clusters considered. Having in mind the inclusion  of light clusters
in other RMF models besides FSU, it is important to study their
behavior at  low densities when matter is clusterized and light
clusters have an important role in the definition of the transport properties
of matter. 

In this paper, we want to apply the generalized RMF (gRMF) formalism to low density clusterized matter, where clusters are treated as new quasiparticles, but with a modified coupling to the $\sigma$ meson field developed in
\cite{PaisPRC97,PaisPRC2019}, using different relativistic mean field models, to understand how they behave. For that matter, we have chosen models with both density-dependent and
density-independent couplings. We consider the four  usual light clusters, that is, $^2$H, $^3$H, $^3$He, and $^4$He, and 
we add, inspired by the work done in Ref. \cite{PaisPRC97},
another cluster, $^6$He, to our calculations. This cluster has been
included in the analysis of the INDRA collaboration \cite{indra}, which has recently published a
new set of equilibrium constants 
\cite{indra} with a controlled bayesian determination of the system density during the expansion, including the possibility of in-medium modifications
 \cite{PaisPRL,PaisJPG}. 
Note that this gRMF approach has been applied recently also to
the description of yields of clusters produced at ternary fission \cite{Natowitz20}.

The aim of the work is to compare different approaches for the RMF
parametrization, the description of the coupling to the meson field,
and the comparison with the recent INDRA data to investigate the influence of correlations on the equation of state. 
This comparison serves as criterion to validate different models for the composition of  subsaturation nuclear matter.

This paper is organized as follows: a brief summary of the formalism applied is given in the next Section, some results are shown in Section \ref{sec:results}, and, finally, in Section \ref{sec:conclusions}, some conclusions are drawn.

\section{Formalism}
\label{sec:1}

We briefly review  the RMF models that will be
considered in the present work, in particular,  FSU2R \cite{FSU2R},
NL3$\omega\rho$ \cite{pais16}, TM1$\omega\rho$ \cite{pais16},  SFHo \cite{sfho}, DDME2 \cite{ddme2}, and DD2 \cite{typel10}. In Table \ref{tab1}, we show some symmetric nuclear matter
properties calculated at saturation density for these models.

\begin{table}
\caption{A few symmetric nuclear matter properties for the models used in this work, calculated at saturation density, $n_0$: the binding energy per particle $B/A$, the incompressibility $K$, the symmetry energy $E_{\rm sym}$, the slope of the symmetry energy $L$, and the nucleon effective mass $M^{*}$. All quantities are in MeV, except for $n_0$ that is given in fm$^{-3}$, and for the effective nucleon mass that is normalized to the nucleon mass $M$.}
\label{tab1}
\begin{tabular}{lllllll}
\hline\noalign{\smallskip}
 Model    &  $n_0$ & $B/A$ &  $K$ & $ E_{\rm sym}$ & $L$ & $M^{*}/M$\\
\noalign{\smallskip}\hline\noalign{\smallskip}
FSU2R & 0.15  & -16.28  & 238  &  30.7  & 47 & 0.59  \\
NL3$\omega\rho$ & 0.148 & -16.24 & 270 & 31.7 & 55 & 0.60 \\
TM1$\omega\rho$ & 0.145 & -16.26   & 280    &  31.6  & 56 & 0.63 \\
SFHo  & 0.158  & -16.19  & 245  &  31.6 & 47 & 0.76  \\
DDME2 & 0.152   & -16.14   & 251  &  32.3 & 51 & 0.57 \\
DD2   & 0.149   & -16.02   & 243  &  32.7 & 58 & 0.56 \\
\noalign{\smallskip}\hline
\end{tabular}
\end{table}

These properties are obtained from the 
Lagrangian density, that describes nucleons with vacuum mass $M$ coupled
to the scalar meson $\sigma$ with mass $m_\sigma$,  the vector
isoscalar meson $\omega$ with mass $m_\omega$ and the vector
isovector meson $\rho$ with mass $m_\rho$,
\begin{equation}
{\cal L}=\sum_{i=p,n} {\cal L}_i + {\cal L}_{\sigma} + {\cal
  L}_{\omega}  + {\cal L}_{\rho} + {\cal L}_{\sigma\omega\rho} \, ,
\label{eq:L}
\end{equation}
where ${\cal L}_i$ is 
$$
{\cal L}_i=\bar \psi_i\left[\gamma_\mu i D^{\mu}-M^*\right]\psi_i \, ,
$$
with
$
i D^{\mu}=i\partial^{\mu}-g_\omega \omega^{\mu}-
\frac{g_{\rho}}{2}  {\boldsymbol\tau} \cdot \mathbf{\rho}^\mu \,, 
$
and the Dirac effective mass,
$
M^*=M-g_\sigma \sigma  \, .
$
The mesonic Lagrangian  densities are given by:
\begin{eqnarray}
{\cal L}_\sigma&=&\frac{1}{2}\left(\partial_{\mu}\sigma\partial^{\mu}\sigma
-m_\sigma^2 \sigma^2 - \frac{1}{3}\kappa \sigma^3 -\frac{1}{12}\lambda\sigma^4\right),\nonumber\\
{\cal L}_\omega&=&-\frac{1}{4}\Omega_{\mu\nu}\Omega^{\mu\nu}+\frac{1}{2}
m_\omega^2 \omega_{\mu}\omega^{\mu} + \frac{\zeta}{4!}\zeta g_\omega^4 (\omega_{\mu}\omega^{\mu})^2, \nonumber \\
{\cal L}_\rho&=&-\frac{1}{4}\mathbf B_{\mu\nu}\cdot\mathbf B^{\mu\nu}+\frac{1}{2}
m_\rho^2 \mathbf \rho_{\mu}\cdot \mathbf \rho^{\mu}+\frac{\xi}{4!} g_\rho^4 (\mathbf\rho_{\mu}\rho^{\mu})^2, \nonumber\\
\end{eqnarray}
where
$\Omega_{\mu\nu}=\partial_{\mu}\omega_{\nu}-\partial_{\nu}\omega_{\mu} ,
\quad \mathbf{B}_{\mu\nu}=\partial_{\mu}\boldsymbol \rho_{\nu}-\partial_{\nu} \boldsymbol \rho_{\mu}
- g_\rho (\boldsymbol{\rho}_\mu \times\boldsymbol{\rho}_\nu)$, and $\boldsymbol \tau$ are the SU(2) isospin matrices.

The Lagrangian density of the models FSU2R,
NL3$\omega\rho$, TM1$\omega\rho$ and  SFHo  includes  non-linear
mesonic  terms, 
which are either self-interaction terms, or terms that mix  the $\sigma, \omega$, and $\mathbf{\rho}$ mesons \cite{sfho}:
\begin{eqnarray}
{\cal L}_{\sigma\omega\rho} &=&  g_{\rho}^2 f (\sigma,\omega_\mu\omega^\mu) 
\boldsymbol{\rho}^{\,\mu} \cdot \boldsymbol{\rho}_{\mu}\;,
\label{FLTL}
\end{eqnarray}
where $f$ is given by
\begin{equation}
f (\sigma, \omega_\mu\omega^\mu) = \sum_{i=1}^{6} a_i \sigma^i 
+ \sum_{j=1}^{3} b_j \left(\omega_{\mu} \omega^{\mu}\right)^{j}\; 
\label{eq:ffun}
\end{equation}
for SFHo, and by
\begin{equation}
f (\sigma, \omega_\mu\omega^\mu) = \Lambda_v g_v^2 \omega_{\mu} \omega^{\mu}\; 
\label{eq:ffun}
\end{equation}
for FSU2R,  NL3$\omega\rho$ and TM1$\omega\rho$.

The models DD2, DDME2 have density-dependent couplings and no
nonlinear mesonic terms.  Their isoscalar couplings of the mesons $i$
to the baryons  are given by  
\begin{eqnarray}
g_{i}(n_B)=g_{i}(n_0)a_i\frac{1+b_i(x+d_i)^2}{1+c_i(x+d_i)^2} \, ,
  i=\sigma, \omega, 
\end{eqnarray}
and the isovector  meson-nucleons coupling by
\begin{eqnarray}
g_{\rho}(n_B)=g_{\rho}(n_0)\exp{[-a_\rho(x-1)]} \, .
\label{grho}
\end{eqnarray}
In the last expressions, $n_0$ is the model-dependent symmetric nuclear saturation density, see Tab.~\ref{tab1}, and $x=n_B/n_0$, with $n_B$ the baryonic density.

In order to include explicitly the light clusters as constituents of nuclear
matter, we consider them as point like particles, 
neglecting intrinsic structures. This approximation is
acceptable at low densities when the volume accessible to each clusters
is much larger than its volume. In RMF approach they will interact
with  the medium through their coupling to the mesons. We add to the Lagrangian density the
cluster  contributions, and  for homogeneous matter each one contributes to the total energy density of
the system with
\begin{eqnarray}
{\cal E}_i=&\dfrac{2S^i+1}{2\pi^2}\int
             k_i^2E_i (f_{i+}(k)+f_{i-}(k))dk_i
  \nonumber\\
&+g_{i\omega}\omega^0 n_i+g_\rho \rho_3^{0}I_3^i n_i \, ,
\end{eqnarray} 
where $E_i=\sqrt{k_i^2+M_i^{*2}}$ is the cluster single-particle
energy, and  $M_i^*$ the
effective mass of cluster $i$. The cluster spin, isospin and density are denoted,
respectively, by
$S^i$, $I_3^i$, and $n_i$.  The distribution functions for the particles and
antiparticles $f_{i\pm}$ are defined as
\begin{eqnarray}
f_{i\pm}&=&\frac{1}{\exp[(E_i\mp\nu_i)/T]+\eta},
\end{eqnarray}
with $\eta=1$ for fermions and $\eta=-1$ for bosons, $T$ the temperature, and
$\nu_i=\mu_i-g_{i\omega}\omega^0-g_{\rho} I_3^i \rho_3^0$. The coupling of the clusters to the
$\omega$-meson is given by $g_{i\omega}=A_i\,g_\omega$, with $A_i$ the cluster mass  number. 

We define the binding energy of the cluster  $i$ in the medium as in \cite{PaisPRC97}
\begin{eqnarray}
B_i=A_i m^*-M_i^* \,, \quad i=d,t,h,\alpha,\, ^6{\rm He} \,, \label{binding}
\end{eqnarray}
with $M_i^*$ given by
\begin{eqnarray}
M_i^*&=&A_i m - g_{i\sigma}\sigma_0 - \left(B_i^0 + \delta B_i\right).
\label{meffi2}
\end{eqnarray}
In this last expression,  $B^0_i$ is the binding energy of the cluster in the
vacuum, which will be fixed  to the  experimental values,   the second
term denotes the coupling of the cluster to the $\sigma$-meson,
and the last term describes a binding energy shift. In
the following, we will write the coupling $g_{i\sigma}$ in terms  of
the coupling of  the nucleon to the $\sigma$-meson, $g_\sigma$, as
$g_{i\sigma}=x_sA_i g_\sigma$, with $x_s$ its fraction, ranging from 0 to 1, to modify (decrease) the nuclear attraction.
For the binding energy shift  $\delta B_i$,  we consider \cite{PaisPRC97,Aymard14,raduta2}
\begin{eqnarray}
\delta
 B_i=\frac{Z_i}{n_0}\left(\epsilon_p^*-m
  n_p^*\right)+\frac{N_i}{n_0}\left(\epsilon_n^*-m n_n^*\right)
 \, ,
\label{deltaB}
\end{eqnarray}
where the gas  energy density  $\epsilon_{j}^*$ and nucleonic density
$n_{j}^*$, ${j}=n,\, p$, are given by \cite{PaisPRC97}
\begin{eqnarray}
\epsilon_{j}^*&=&\frac{1}{\pi^2}\int_0^{k_{F_{j}}(\rm gas)} k^2 E_{j} (f_{{j}+}(k)+f_{{j}-}(k)) dk \\
n_{j}^* &=&\frac{1}{\pi^2}\int_{0}^{k_{F_{j}}(\rm gas)}  k^2 (f_{{j}+}(k)+f_{{j}-}(k)) dk \,,
\end{eqnarray}
with $ k_{F_{j}}({\rm gas})= \left(3\pi^2 n_{j}\right)^{1/3}$ defined  using the zero temperature relation between density and Fermi momentum.  This was proposed in \cite{PaisPRC97}  as an effective way of implementing Pauli blocking in a range of temperatures  for which the Fermi distribution is close to a  step function.
These two quantities,  $\epsilon_{j}^*$ and $n_{j}^*$, define the energy density and
the density of nucleon $j$ associated to the levels below the Fermi momentum,
 $ k_{F_{j}}({\rm gas})$, of the $T=0$ nucleonic density $n_{j}$.  
The term $\delta
 B_i$ can be identified as  an excluded volume mechanism in the
 Thomas-Fermi approximation.
 The gRMF model is able to describe the occurrence of clusters in nuclear matter and density modifications.
 The parameters are introduced in an empirical way, and it is our goal to discuss different choices for these  parameters, in particular the coupling of clusters $x_s$ to the $\sigma$-meson. 
 There are also alternatives to describe density effects such as the excluded volume concept, see \cite{fischer13}, which also uses empirical parameters. A microscopic, quantum statistical approach can be given, see \cite{Roepke15} and references therein, which, however, is not  simple to be used for practical applications.

\section{Results}
\label{sec:results}

\begin{figure}
   \includegraphics[width=0.99\linewidth]{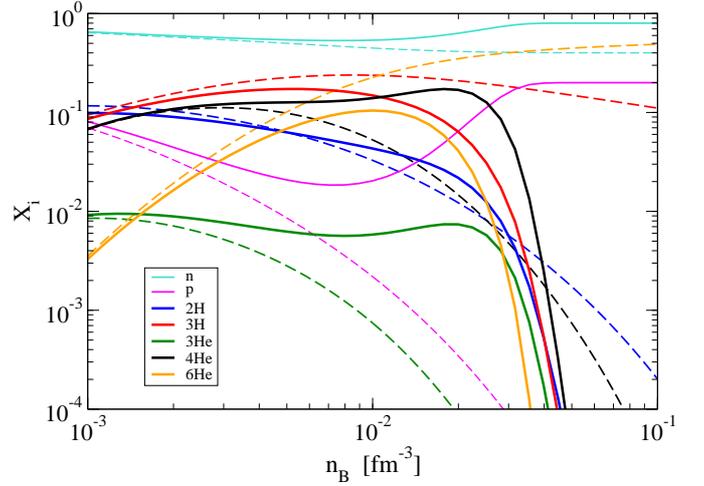} 
 \caption{(Color online) The abundances (mass fraction) of the stable isotopes $n, p, d, t, h, \alpha$, and $^6$He considered as a function of the density for $T=5$ MeV and a fixed
   proton fraction of 0.2.  The NSE (dashed) is compared to a QS calculation (full lines), see text.}
\label{fig0}
\end{figure}

Before comparing different gRMF models, we give a general discussion
of the problem to include correlations and cluster formation in the EoS. 
In Fig.~\ref{fig0}, the mass fraction $X_i=A_i n_i/n_B$ for the stable isotopes $i=n, p, d, t, h, \alpha$, and $^6$He
is shown for $T=5$ MeV, $y_{p}=0.2$ in the baryon number density region $0.001\, {\rm fm}^{-3} < n_B < 0.1\, {\rm fm}^{-3}$; $y_p$ is the fixed global proton fraction, and $n_i$ the particle number density. We have $\sum_i X_i = 1$.
The calculation of the composition according to the 
simple model of nuclear statistical equilibrium (NSE) neglecting all interactions
between the constituents, and considering only the ground states of the stable isotopes,
is compared to a quantum statistical (QS) calculation taking in-medium effects into account, 
in particular Pauli blocking and the quasiparticle shift taken as DD2-RMF, see \cite{Roepke15}.
In the low-density region, the interaction between the constituents of nuclear matter can be neglected
so that this limiting region is appropriately described by the NSE. A small difference is 
observed for the deuteron mass fraction owing to the virial limit which is correctly described by the QS approach.
The account of excited states implements also the account of scattering states in the virial EoS \cite{virial1,virial2} which leads to a
significant contribution for the deuteron fraction because of its small binding energy. As seen from Fig.~\ref{fig0}, in-medium effects become appreciable above $n_B=10^{-3}$ fm$^{-3}$.

The most striking effect is the suppression of bound state abundances because of Pauli blocking so that
near the saturation density these clusters nearly disappear, and we obtain a Fermi liquid of neutrons and protons.
These are treated as quasiparticles containing a mean-field energy shift. 
Different versions of these RMF approaches are presented and compared in this work. 
We have to interpolate between two limiting cases, the virial expansion in the low-density
limit and the RMF approach near the saturation density. In particular, the virial EoS
depends only on the experimentally determined binding energies and scattering phase shifts
\cite{virial1,virial2}, and provides at finite temperature the correct zero density limit.
 The exact microscopic description of the intermediate region is a difficult many-particle problem.
Continuum correlations and higher order clustering in a dense medium is hard to calculate within the QS approach.
Interpolations are of interest which may be probed by laboratory experiments as shown in this paper.

An interesting issue seen also in Fig.~\ref{fig0} is that within the NSE calculation with a given set of isotopes, 
the neutron-rich clusters become dominant 
at increasing density, because the proton fraction is small so that exotic nuclei like $^6$He are most abundant near the saturation density.
We can extend the NSE including resonant states such as $^4$H and $^5$He. In a recent publication by Yudin et al. \cite{yudin19}, it was claimed that these exotic nuclei may be of importance in stellar matter, in particular with respect to the neutrino opacity.
However, as shown in Refs.~\cite{Roepke20,Fischer20}, within a more systematic approach the contribution of the resonant states should be expressed in terms of the scattering phase shifts so that their contribution is strongly reduced. Therefore, in this paper, we restrict our calculations only to $^6$He, which is also measured in the INDRA experiment. 
Within a more exhaustive investigation one can also search for $^8$He and other clusters, but it is expected that their mass fractions are very small.
We will not go in more details here, but discuss what can be learned from laboratory experiments 
to derive adequate interpolation expressions to infer the composition in the whole subnuclear density region.

\begin{figure}
   \includegraphics[width=0.99\linewidth]{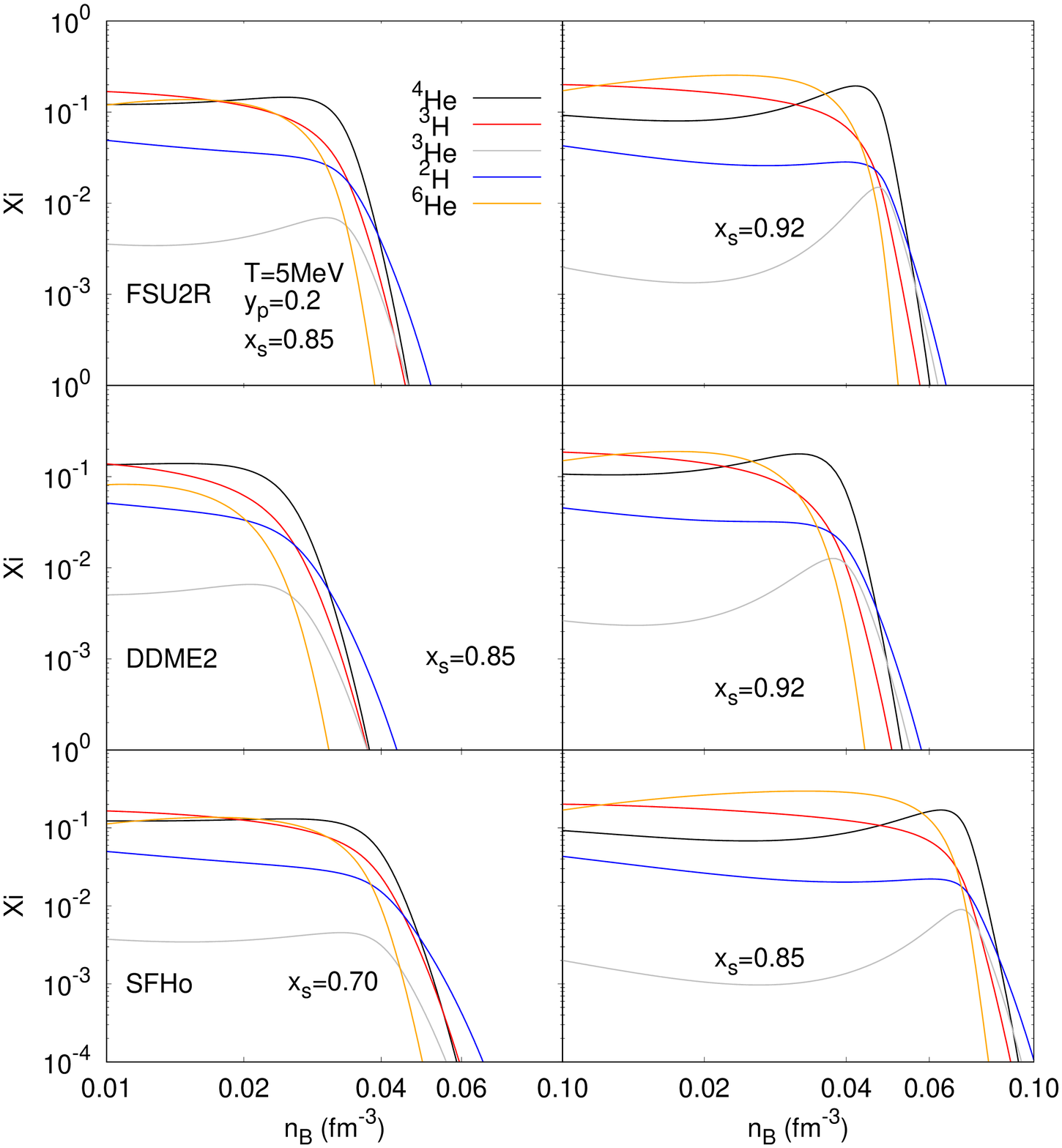} 
 \caption{(Color online) The abundances of all the clusters
   considered as a function of the density for $T=5$ MeV and a fixed
   proton fraction of 0.2 for FSU2R (top), DDME2 (middle), and SFHo (bottom) obtained
   with $x_s=0.85$ (left) and   $x_s=0.92$
   (right). For the SFHo model, the scalar cluster-meson coupling used
   was 0.7 and 0.85. } 
\label{fig1}
\end{figure}

To this aim, in the present section, we compare the distribution of light clusters
at low densities obtained with the models  FSU2R,
NL3$\omega\rho$, TM1$\omega\rho$,  SFHo, DDME2, and DD2, and we discuss
how well these models describe the equilibrium constants
determined from the INDRA \cite{indra,PaisPRL,PaisJPG} data.

As referred before, in \cite{PaisPRC97} it was shown that taking the
universal coupling   $x_s=0.85\pm0.05$ of the
$\sigma$-meson to the light
clusters  $d,\, t,\, h,\, \alpha$  would  describe well the equilibrium constants
determined from the NIMROD data \cite{qin12}. These data, however,
have  been analyzed making the hypothesis that the system volume
can be obtained assuming an ideal gas equation of state for the clusters.
 Recently, the INDRA collaboration \cite{indra} has
performed a similar experiment with the heavier system  Xe-Sn, where
different  tin and xenon isotopes were used, and considered in the analysis 
also the $^6$He  cluster, besides
the light clusters $d,\, t,\, h,\, \alpha$.
The  Bayesian analysis of these data performed in \cite{PaisPRL,PaisJPG},
allowing possible in-medium corrections in the determination of the system volume,  has  shown that a larger value $x_s=0.92\pm0.02$
should  be taken for the cluster $\sigma$-meson coupling.
However, this value is obtained with a specific version of the RMF model, namely FSU, and 
it could be model dependent. For this reason,  
in  the following figures, we will consider different values of $x_s$ for the
$\sigma$  meson-cluster coupling within  the RMF models introduced above.

In Fig.~\ref{fig1}, we show the mass fractions of all the clusters
considered  as a function of the density for a temperature of 5 MeV
and a fixed proton fraction  $y_{p}=0.2$ .
We have chosen three models: two with nonlinear mesonic terms and constant couplings (FSU2R and SFHo), and another with density-dependent couplings (DDME2). We have taken the 
scalar cluster-meson coupling fraction as 0.85 and 0.92 for both FSU2R
and DDME2. For SFHo, we have considered a smaller fraction, 0.7 and 0.85,
because taking a larger fraction, e.g. 0.92, would make the dissolution of
the clusters occur at a much larger density, as we will see in the
following. Moreover, a smaller $x_s$ is necessary to reproduce the virial
EoS at low densities, and to fit the equilibrium constants deduced from
the INDRA data as will be discussed later.
We can see that  below  $n_B \approx 3\times 10^{-2}$  fm$^{-3}$, the
most abundant clusters are $^3$H, $^4$He, and  $^6$He. This is because the clusters
$^6$He and $^3$H are the neutron-richest ones,  and  we are considering
asymmetric neutron rich matter,  and $^4$He is the most
bound one. The heaviest cluster $^6$He  is the first to dissolve while
the lightest one $^2$H is the last one. For
FSU2R and SFHo,  $^6$He is the most abundant cluster in a short range of
densities taking, respectively,  $x_s=0.85$ and 0.7. It is striking that  for
$x_s=0.92$ (0.85) for FSU2R (SFHo),
$^6$He becomes the most abundant in the range 0.01 to 0.04
fm$^{-3}$. 
Within  DDME2, the clusters dissolve at the smallest densities, 
and  $^6$He is the most  abundant cluster in a short
range of densities  only when the larger value of $x_s$ is considered.
For $x_s$=0.85, the dissolution density range for all clusters is
$\sim 4 - 6 \times 10^{-2}$ fm$^{-3}$ for FSU2R and DDME2.  It is interesting to see that for the
RMF  with non-linear terms, such as FSU2R and SFHo,  the fraction of $^6$He
 clusters becomes quite large just before the binding energy of the
 cluster goes to
 zero, followed by a steep decrease of the abundance of this
 cluster. In the case of SFHo and $x_s$=0.85,  we can even
 identify a first order phase transition  at this density.

\begin{figure}
		\includegraphics[width=0.99\linewidth]{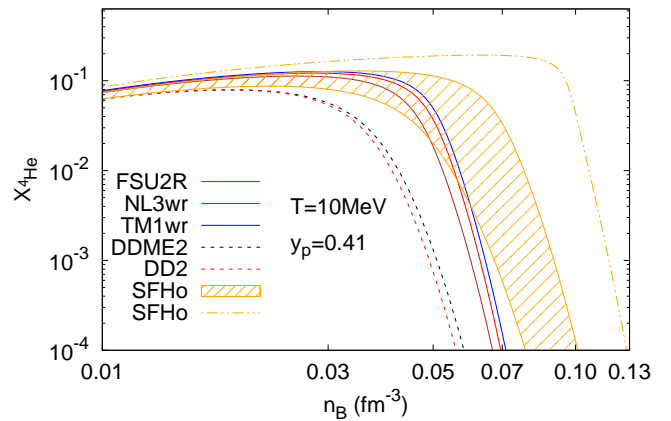} 
	\caption{(Color online) The mass fraction of the $\alpha$ cluster as a function of the density for all the models considered, and $T=10$ MeV. The scalar cluster-meson coupling fraction is set to $x_s=0.85$, except for SFHo where the results are shown for $x_s=0.7\pm 0.05$ (wide hashed band) and for $x_s=0.85$ (dash-dotted-line). The proton fraction is fixed at 0.41. } 
	\label{fig4}
\end{figure}

In Fig.~\ref{fig4}, we compare the $\alpha$-particle mass distributions
obtained for the models FSU2R, NL3$\omega\rho$, TM1$\omega\rho$,
SFHo, DD2 and DDME2, considering a temperature of $T=10$ MeV, and proton
fraction of $y_{p}=0.41$. For all the models shown, the scalar cluster-meson coupling
fraction is set to $x_s$=0.85. For SFHo, we have also plotted the abundances with $x_s=0.70\pm 0.05$, represented by the hashed region. The differences reflect the properties of these
models at subsaturation densities. Below 0.02 fm$^{-3}$,
i.e. not far from the range of densities where the virial EoS is
valid,  all models
give similar results. Above 0.03 fm$^{-3}$, the models start to
differ, with DD2 and DDME2 predicting the smallest dissolution density,
slightly below  0.05 fm$^{-3}$, and SFHo the largest one, $\approx 0.1$
fm$^{-3}$. SFHo is, in fact, a special case because all the other models predict
dissolution densities in a narrow range of  $\sim 0.05-0.06$
fm$^{-3}$. With SFHo, we get a similar result, if the $x_s$ is reduced to
$\approx 0.65-0.7$.  It is expectable that under the conditions where
light clusters play an important role, the predictions obtained with
SFHo will differ from the ones obtained with any of the other five
models. This could be expected from the results of Ref.~\cite{olfa20}
  where the spinodal sections obtained for SFHo extend to a much larger ($n_p,\, n_n$) phase-space region than FSU2R, TM1$\omega\rho$ or DDME2.

\begin{figure*}
		\includegraphics[width=1\linewidth]{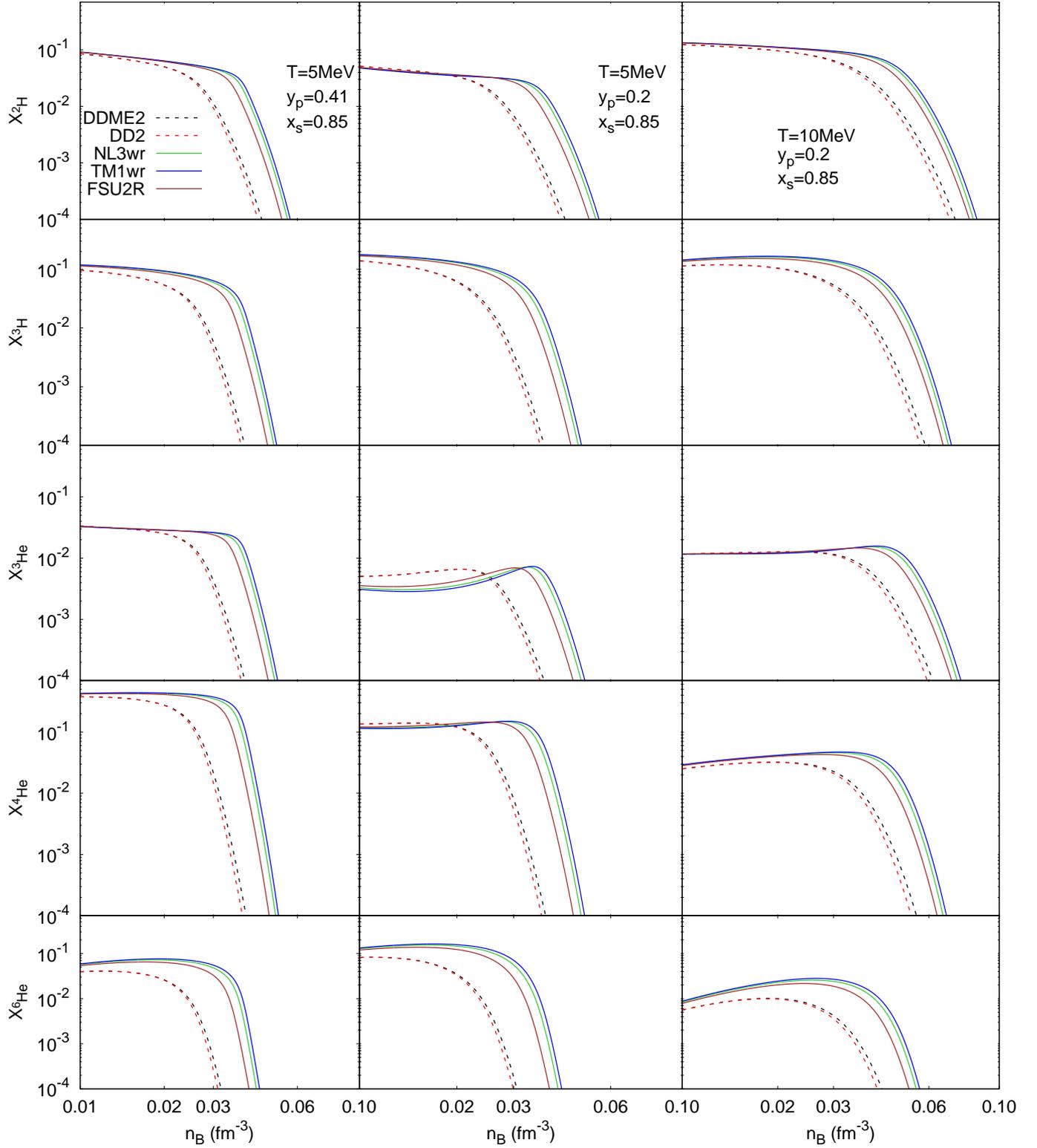} \\ 
	\caption{ (Color online) The mass abundances of all the clusters considered as a function of the density  for  the models FSU2R, NL3$\omega\rho$, TM1$\omega\rho$,
SFHo, DD2 and DDME2  and $T=5$ MeV and $y_{p}=0.41$ (left column),
$T=5$ MeV and  $y_{p}=0.2$ (middle column), and $T=10$ MeV and $y_{p}=0.2$ (right column). In all calculations, the scalar cluster-meson coupling is set to $x_s=0.85$.}
	\label{fig3}
\end{figure*}

In Fig.~\ref{fig3}, we plot, for comparison, the mass distributions of the five clusters considered, calculated within the models FSU2R, NL3$\omega\rho$, TM1$\omega\rho$, SFHo, DD2 and DDME2, for two temperatures, $T= 5$, and 10 MeV, and two proton fractions $y_p=$ 0.2 and 0.41. The temperatures chosen are typical in proto-neutron stars, and the proton fractions  reflect two different stages of the star evolution.
Some general comments are in order: (i) 
 In average, models with density-dependent couplings give different
  fractions from  the others if the same value of  $x_s$ is chosen. They generally predict the cluster dissolution at smaller
  densities and smaller particle fractions at low densities for the
  neutron rich clusters; (ii) In the low-density range shown, the neutron rich clusters, tritium
  and $^ 6$He, are the most abundant clusters for $y_{p}=0.2$ and $T=5$ MeV. At $T=10$
  MeV, this is still true for the tritium, the one with the  smallest
  mass; (iii) Models with non-linear mesonic terms show a steeper behavior
  close to the dissolution density.

We now turn to examine how the differences observed in the models reflect in the 
predictions for the equilibrium constants, which are the quantities determined from the experimental
data \cite{qin12,indra}.
The equilibrium constants are defined as
the ratio
$$
K_{ci}=\frac{n_i}{n_n^{N_i} n_p^{Z_i}},
$$
where  $n_i$ is the density of cluster $i$, $n_n$ and $n_p$
are,  respectively,  the density of free neutrons and protons, and
$Z_i$, $N_i$ are the number of  protons and neutrons in cluster $i$.

\begin{figure*}
		\includegraphics[width=0.9\linewidth]{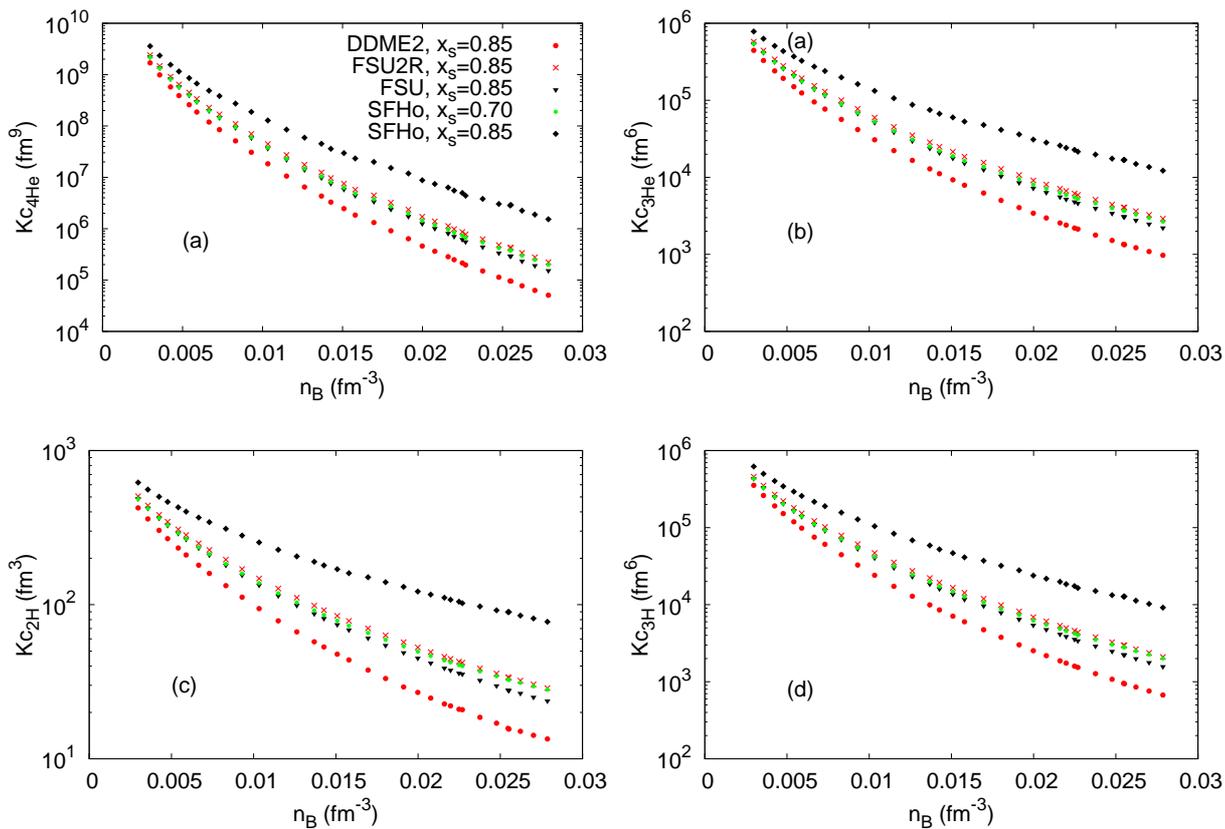} 
	\caption{(Color online) The chemical equilibrium constants as a function of the density from a calculation where we consider homogeneous matter with four light clusters for the
DDME2 (red circles), FSU2R (crosses), and FSU (black triangles)
  models, calculated at the experimental values ($T$, $n_{B_{\rm
  exp}}$, $y_{{p}_{\rm exp}}=0.41$) proposed in
  Ref.~\cite{qin12}, and considering the cluster coupling as
  $x_{s} = 0.85$. Also shown are the results for SFHo with $x_{s} =
  0.7$ (yellow circles) and $x_{s} = 0.85$ (black diamonds). }
	\label{fig5}
\end{figure*}

In Fig.~\ref{fig5}, we plot the equilibrium constants obtained 
with the different models on typical ($n_B-T$) trajectories that can be explored in heavy ion collisions.
The choice of the temperature value, at each density point, is the one estimated in Ref.~\cite{qin12}, and the proton fraction is fixed to $y_{p}=0.41$ at each point.
The volume estimation in that paper is not fully realistic, since it was made in the simplifying assumption of an ideal cluster 
gas. However, many different theoretical calculations
\cite{hempel2015} were produced assuming the $(T,\, n_B,\, y_{p})$ correlation 
of Ref.~\cite{qin12}, therefore this particular choice is useful to
assess the model dependence of the calculations.
In this Figure, the sensitivity of the chemical constants to the scalar cluster-meson coupling for one representative model, SFHo, is shown. We can see that the effect previously observed in Fig.~\ref{fig4}, namely the positive correlation between the value of $x_s$, here reflects into  higher values for the equilibrium constants when $x_s$ is increased. This effect is sizeable and potentially 
bigger than the experimental error bars on equilibrium constants,
meaning that a comparison with experimental data, within a given
model, allows predicting the dissolution density of clusters in dense
matter.  The results  of  three different
models DDME2, FSU2R and SFHo, using the same value for 
that coupling, fixed to   $x_s=0.85$, as proposed in \cite{PaisPRC97}, are also displayed in  this figure, and they show the model dependence of the equilibrium constant prediction.
Models TM1$\omega\rho$ and  NL3$\omega\rho$ have a behavior very close to FSU2R, and are not represented. The difference between the predictions is a measure 
of the model dependence of the calculation. We can see that the chemical constants are smaller  for DDME2  reflecting  the fact discussed  before that the clusters dissolve at smaller densities within this model.  We can,  however, say that  the behavior of  FSU2R and DDME2 models is  similar to the one
  obtained in \cite{PaisPRC97} within the model
FSU,  probably due to the fact that these models have
similar properties at subsaturation densities.

\begin{figure*}
		 \includegraphics[width=0.3\linewidth]{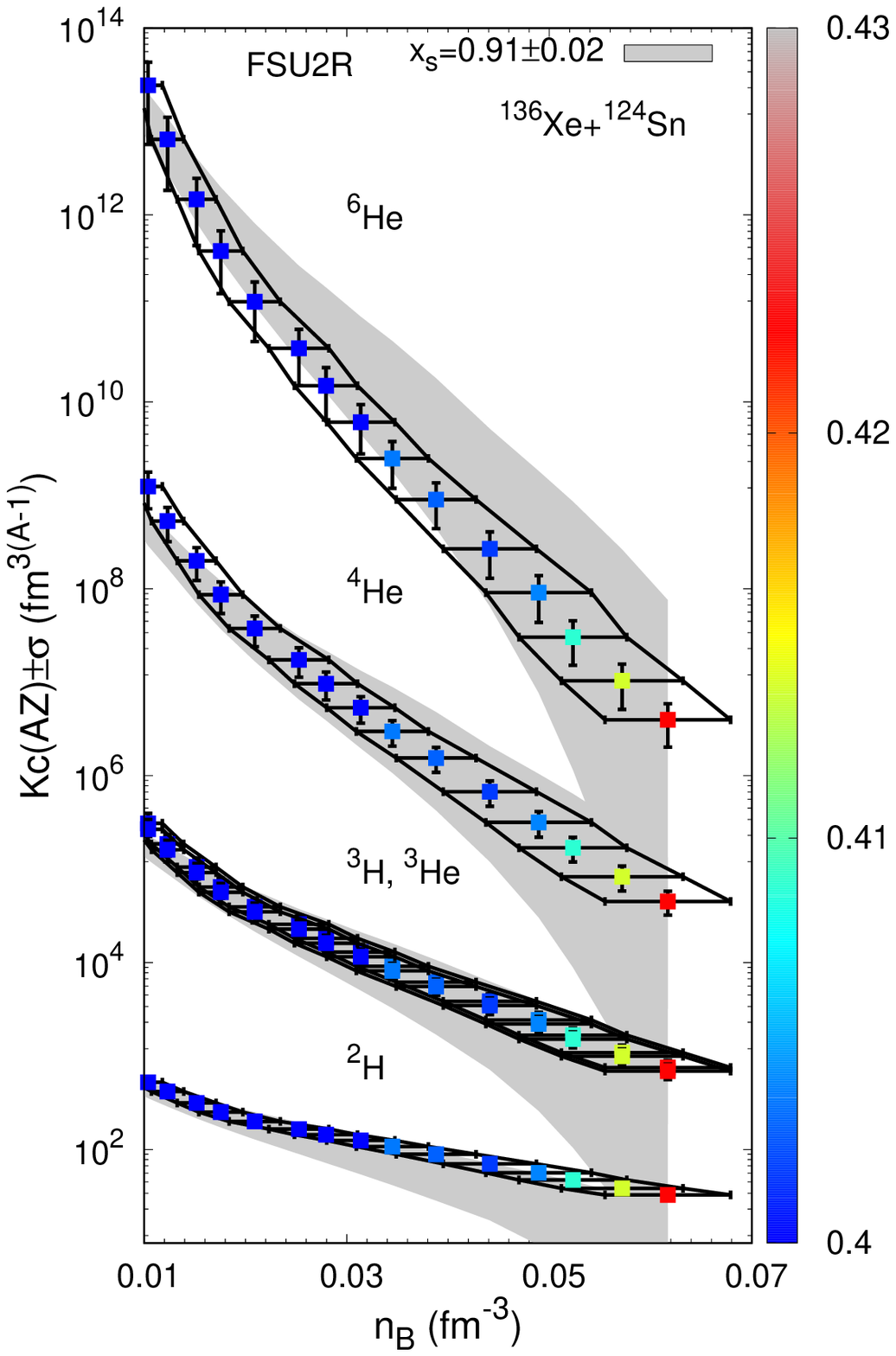} 
		 \includegraphics[width=0.3\linewidth]{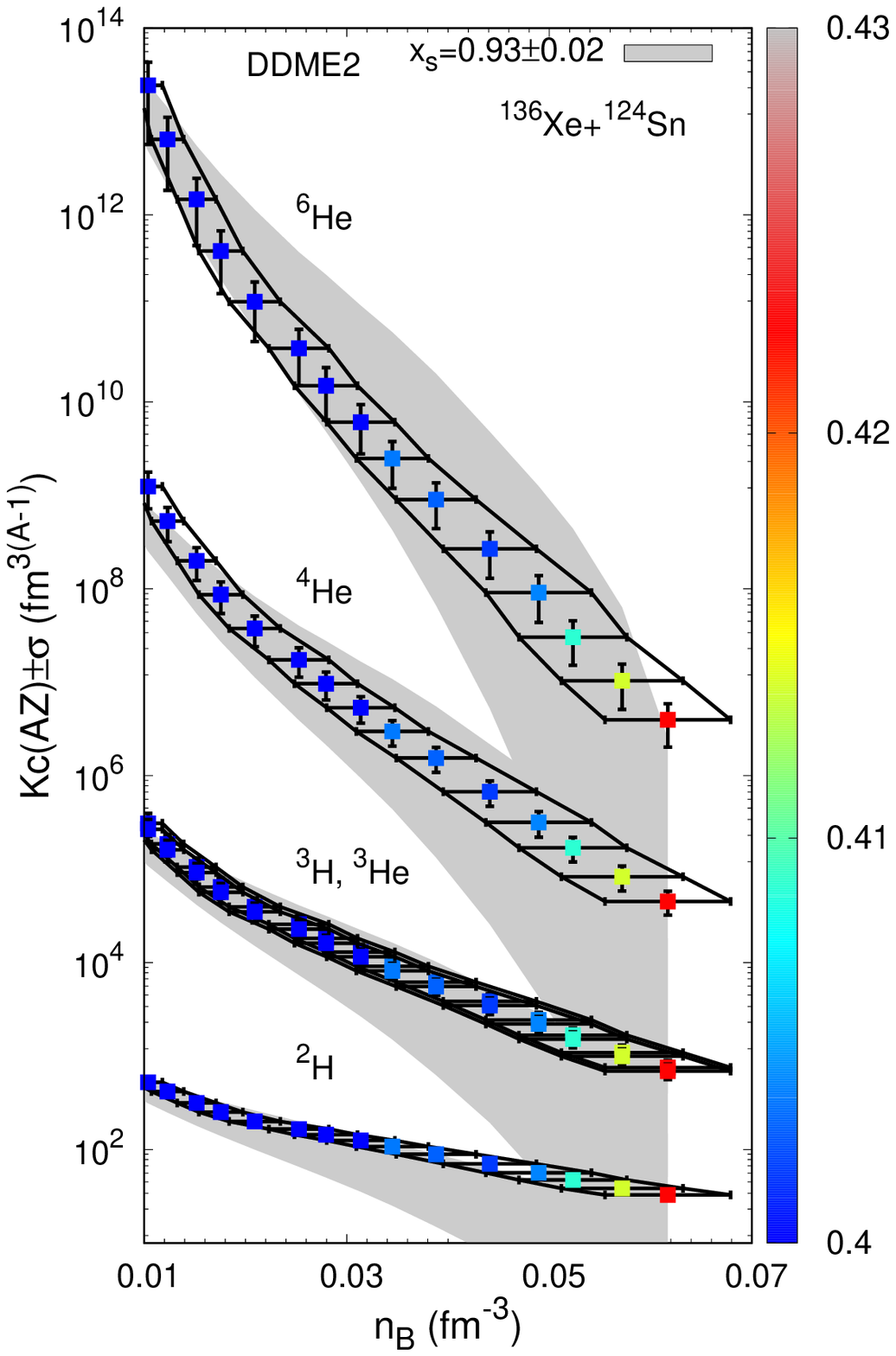}
		 \includegraphics[width=0.3\linewidth]{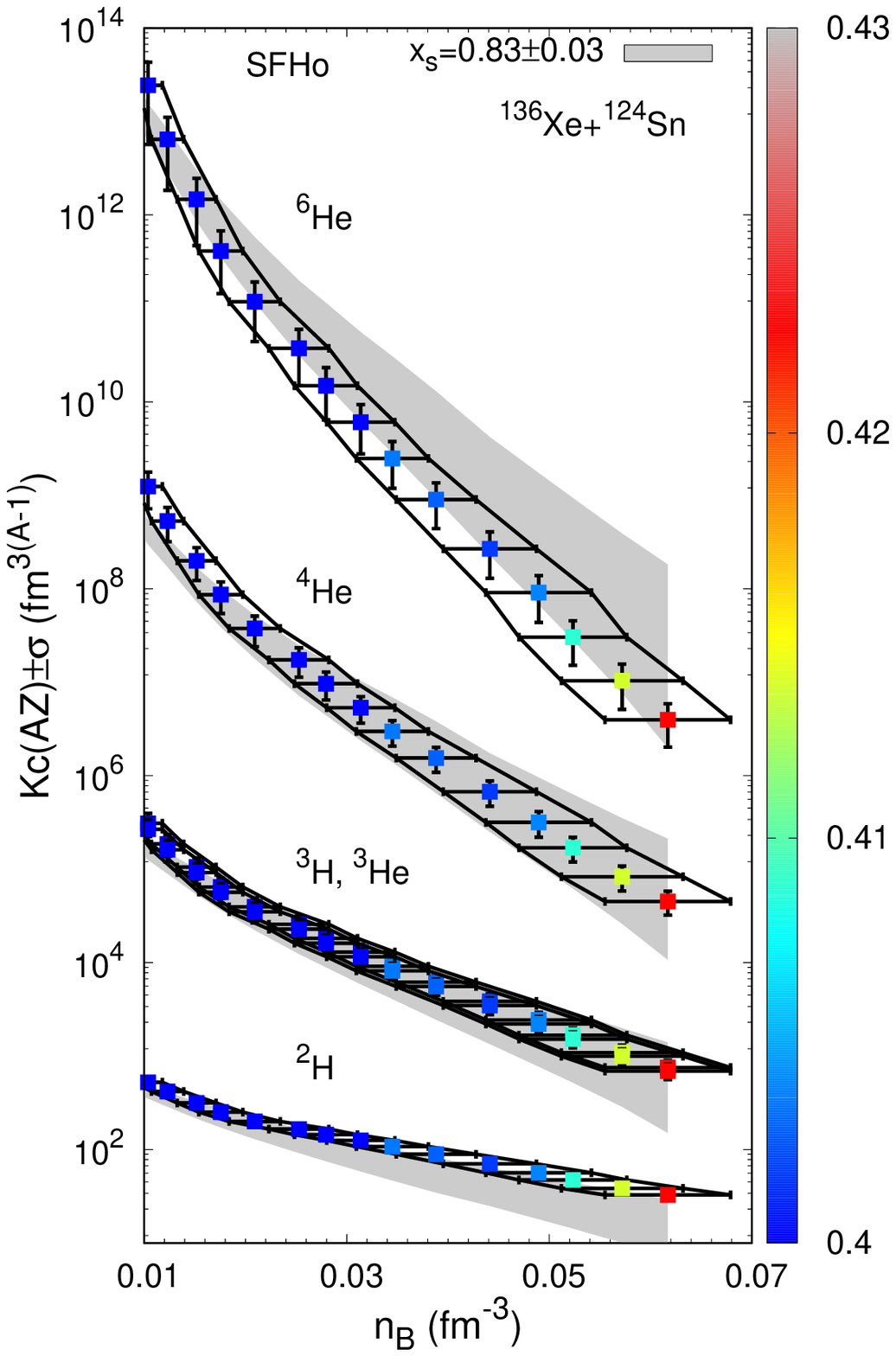} 
	\caption{(Color online) The equilibrium constants as a function of the
          density. The full lines represent the experimental
          results of the INDRA collaboration, with $1\sigma$
          uncertainty. The grey bands are the equilibrium constants
          from a calculation ~\cite{PaisPRC2019} where we consider
          homogeneous matter with five light clusters for the FSU2R
          EoS (left), the DDME2 EoS (middle) and SFHo  EoS (right),
          calculated at the average value of ($T$, $n_{B_{\rm exp}}$,
          $y_{{p}_{\rm exp}}$), and considering cluster couplings in
          the range of $x_{s} = 0.91\pm 0.02$ (FSU2R),  $x_{s} = 0.93\pm 0.02$ (DDME2)  and
          $x_{s} = 0.83\pm 0.03$ (SFHo). The color code represents the global proton fraction. }
	\label{fig7}
\end{figure*}

\begin{figure}
   \includegraphics[width=0.99\linewidth]{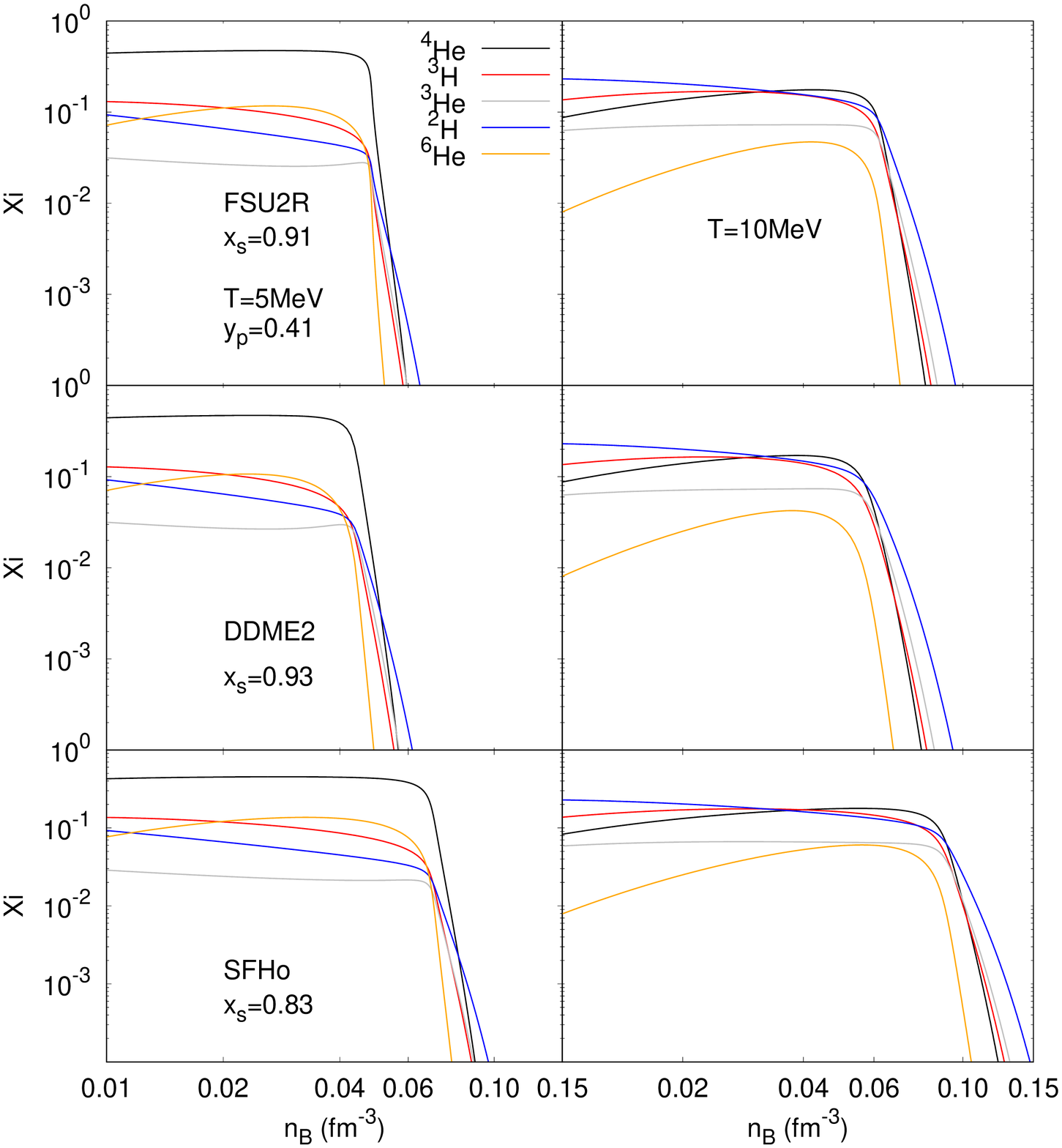} 
 \caption{(Color online) The abundances of all the clusters
   considered as a function of the density for $T=5$ MeV (left) and $T=10$ MeV (right) and a fixed
   proton fraction of 0.41 for FSU2R with $x_s=0.91$ (top), DDME2 with $x_s=0.93$ (middle), and SFHo with $x_s=0.83$ (bottom). } 
\label{fig6}
\end{figure}

As expected, the coupling
$x_s$ is model-dependent, and it should be fitted to some
experimental data or ab-initio calculation. In the following, we determine
for  each model the range of 
$x_s$ values that  best reproduces the INDRA equilibrium
constants for the different clusters as
calculated in  \cite{PaisPRL}.  
 In Fig.~\ref{fig7},  we show how the chemical
 equilibrium constants calculated within models FSU2R, DDME2 and SFHo
 with the best $x_s$ values
 compare with the ones extracted from the experimental data obtained
 by the INDRA collaboration.  
We have obtained for FSU2R
$x_s=0.91\pm 0.02$, for DDME2  $x_s=0.93\pm 0.02$, and for
SFHo  $x_s=0.83\pm 0.03$. It is observed that the quality of
the fit depends slightly on
the size of the clusters, and this deserves an investigation in a
future work.
Let us recall that  previously, in Refs. \cite{PaisPRL,PaisJPG}, the
 authors performed an analysis of the experimental data  within the
 FSU model, and in order to
 reproduce data, it was necessary to take $x_s=0.92\pm 0.02$, a
 result very close to the one obtained with  FSU2R and DDME2.  For
 SFHo, and as we saw before, we need a smaller coupling $x_s$ to fit this data.

\begin{figure}
   \includegraphics[width=0.99\linewidth]{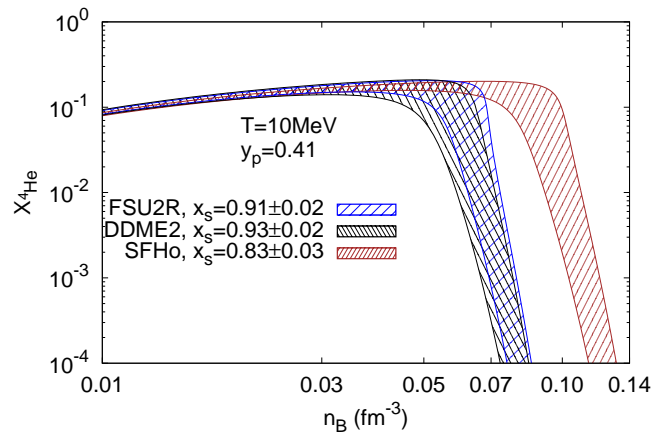} 
 \caption{(Color online) The mass fraction of the $\alpha-$particle
   as a function of the density for $T=10$ MeV and a fixed
   proton fraction of 0.41 for FSU2R with $x_s=0.91\pm 0.02$, DDME2 with $x_s=0.93\pm 0.02$, and SFHo with $x_s=0.83\pm 0.03$. } 
\label{fig8}
\end{figure}

Choosing  the scalar cluster-meson coupling ratio that best fits
  the INDRA data, we calculate the clusters abundances for FSU2R,
  DDME2 and SFHo, for $y_{p}=0.41$ and $T=5$ and 10 MeV, see
  Fig.~\ref{fig6}.  
All models predict similar abundances of all the clusters considered up
to a density $\approx 0.05$ fm$^{-3}$ for $T=5$ MeV and $\approx 0.06$
fm$^{-3}$ for $T=10$ MeV. This result is very interesting: in fact, as
shown in \cite{PaisJPG}, the INDRA data explore densities up to
$\approx 0.06$ fm$^{-3}$, however these larger densities are attained
at a temperature $\approx 9$ MeV.  Matter at  $T=5$ MeV
corresponds to  densities below  $\approx 0.02$
fm$^{-3}$. We may, therefore, expect that a fit to the INDRA data is
giving information on the abundances of light clusters corresponding
to a pair ($T$, $n_B$). Although the proton fraction is also changing along
with $T$ and $n_B$, it takes  values in a very narrow range, 0.39-0.42, very close to $y_{p}=0.41$ used to calculate the cluster abundances in Fig.~\ref{fig6}.

Models FSU2R and DDME2 show very similar fractions also above $\approx 0.05$
  fm$^{-3}$, in particular,  at the maximum of
  the distributions and at the dissolution density,  which we define
  as the density  above which the fractions are below
  10$^{-4}$. However, SFHo  predicts dissolution densities $\sim 30\%$ larger than the other two
models. 

Having this in mind, we plot in Fig.~\ref{fig8} the mass fraction of
the $\alpha-$particle as a function of the density for the three
models previously considered, and a temperature of 10 MeV and a fixed
proton fraction of 0.41. For each model, we choose the range of the
$\sigma-$coupling that best fits the INDRA data. We confirm that
the $\alpha$-abundances predicted by the three models coincide up to $\approx 0.06$
  fm$^{-3}$. Moreover, for FSU2R and DDME2 we do have a  complete
  superposition of the bands, indicating a similar prediction for the
  dissolution density. SFHo, however, shows a higher dissolution
  density, $\sim 30\%$ larger.

 The present results seem to indicate that a good reproduction of the
equilibrium constants obtained from the experimental data could imply a
unique prediction for the cluster abundances, and, in particular, of
the dissolution density  only if we could have some extra experimental constraints 
at a slightly higher density.

\section{Conclusions} \label{sec:conclusions}

We have analyzed the appearance of light clusters in warm
non-homogeneous matter at densities below saturation density in
the framework  of  RMF models. We used six models that  properly describe  nuclear
matter properties, and predict stars with more than two
solar masses, two of which with density-dependent
couplings, and the other four with non-linear mesonic terms. Light clusters
were included as point-like particles that are affected by the medium
through their  couplings to the mesons.  For these couplings, we have
considered: (a) the results of \cite{PaisPRC97}, where, for the
$\sigma$-meson coupling, a universal coupling proportional to $A_i x_s
g_\sigma$, with  $x_s$ to be fixed on experimental data,  was proposed; (b) the
couplings  determined in \cite{PaisPRL,PaisJPG} extracted from the
INDRA \cite{indra} experimental data.

 Except for the model SFHo, we have found that different models predict similar abundances of
clusters. Overall, for the density-dependent models we have obtained
15\% to 20\% smaller dissolution densities, but  far from the
dissolution density, the abundances are similar with respect to the non-linear models. For SFHo,  taking the same
scalar cluster-meson coupling, the dissolution  densities are
approximately the double, and the cluster abundances are larger. It is, therefore, expectable that simulations that use SFHo to describe supernova explosions or binary NS mergers will have larger
contributions of light clusters. In order to reproduce the equilibrium
contants obtained from heavy ion collisions,  a smaller coupling of the
light clusters to the $\sigma$-meson has to be considered. We
  conclude that the clusterization effect, in particular the amount
  and the chemical composition of clusters, depends on the behavior of the model
  in the corresponding density range. Taking universal couplings for
the clusters highlights the differences. The present heavy ion
constraints are not enough to distinguish between models like DDME2
and FSU2R, but clearly shows that SFHo requires a different treatment.

 In the present comparison, we have considered besides the lighter
  clusters  $d,\, t,\, h,$ and $\alpha$, also the heavier cluster
  $^6$He. In asymmetric matter, it was shown that the contribution of
  this cluster is quite important in a range of densities not far
  from the dissolution density. A discussion of the role of
  heavier clusters at the densities and  temperatures studied in the
  present work has been presented in \cite{PaisPRC2019}.  Moreover, we believe there is a need of experimental measurements for heavier clusters in order to discriminate the different models.

The gRMF formalism presented here allows to take cluster formation into account for
hot and dense nuclear matter, in particular stellar matter.
For the contribution of nucleon quasiparticles ($n, p$) different parametrizations
within the RMF are possible. We considered several models, and some of them were calibrated to the INDRA data, namely FSU2R, DDME2, and SFHo. The coupling parameter
$x_{s}$ for the interaction with the $\sigma$ field can be introduced as a global
quantity for all clusters. It determines the density where the
respective clusters are dissolved. We have shown that if $x_{s}$  is fitted
  to equilibrium constants determined from experimental data,
  different models predict similar abundances up to the densities and
  temperatures explored by INDRA data. 
The dissolution densities, however, differ: while two of the models,
FSU2R and DDME2 predict similar behavior at dissolution, the third
model, SFHo, gives  dissolution densities that are at least 30\% larger.
 In the  future, a more
  careful analysis will  be undertaken using statistical methods to
  extract these quantities. 
Besides, a microscopic approach to this coupling parameter may show a dependence on the respective nucleus, as well  as on thermodynamic parameters, like the temperature. This
may indicate that the model applied in the present study needs to take
these dependences into account. This point is left for future developments.

\begin{acknowledgement}

We thank R. Bougault for useful discussions. This work was supported by the FCT (Portugal) Projects No.  UID/FIS/04564/2019 and UID/FIS/04564/2020, and POCI-01-0145-FEDER-029912, by PHAROS COST Action CA16214, and by the German Research Foundation (DFG), Grant \# RO905/38-1. H.P. acknowledges the grant CEECIND/03092/2017 (FCT, Portugal). 

\end{acknowledgement}


\begin{thebibliography}{99}

\bibitem{barranco80} M. Barranco and J. Robert Buchler,  Phys. Rev. C {\bf 22}, (1980) 1729.
\bibitem{muller95} H. M\"uller and B. D. Serot, Phys. Rev. C {\bf 52}, (1995) 2072.
\bibitem{borderie19} B. Borderie and J.D. Frankland, Prog. Part. Nuc. Phys. {\bf 105}, (2019) 82.
\bibitem{arcones08} A. Arcones, G. Martinez-Pinedo, E. O'Connor,
  A. Schwenk, H.-T.Janka, C. J. Horowitz, and K. Langanke, Phys. Rev. C {\bf 78}, (2008) 015806.
\bibitem{ropke08} K. Sumiyoshi and G. R\"opke, Phys. Rev. C {\bf 77}, (2008) 055804.
\bibitem{fischer13} T. Fischer, M. Hempel, I. Sagert, Y. Suwa, and J. Schaffner-Bielich, Eur. Phys. J. A {\bf 50}, (2014) 46.
\bibitem{furusawa13} S. Furusawa, H. Nagakura, K. Sumiyoshi, and S. Yamada, Astrophys. J. {\bf 774}, (2013) 78.
\bibitem{furusawa17} S. Furusawa, K. Sumiyoshi, S. Yamada, and H. Suzuki, Nucl. Phys. A {\bf 957}, (2017) 188.
\bibitem{bauswein13} A. Bauswein, S. Goriely, and H.-T. Janka, Astrophys. J. {\bf 773}, (2013) 78.
\bibitem{fernandez13}  R. Fernandez,  B. D. Metzger,   Mon. Not. Roy. Astro. Soc. {\bf 435},  (2013) 502.
\bibitem{just14} O. Just, A. Bauswein, R. A. Pulpillo, S. Goriely, H.-T. Janka, Mon. Not. Roy. Astro. Soc. {\bf 448}, (2014) 541.
\bibitem{ravenhall83} D. Ravenhall, C. Pethick, and J. Wilson, Phys. Rev. Lett. {\bf 50}, (1983) 2066.
\bibitem{schneider13} A. S. Schneider, C. J. Horowitz, J. Hughto, and D. K. Berry, Phys. Rev. C {\bf 88}, (2013) 065807.
\bibitem{horowitz14} C. J. Horowitz, D. K.  Berry, C. M.  Briggs, M. E.
  Caplan, A. Cumming, and A. S. Schneider, Phys. Rev. Lett. {\bf 114}, (2014) 031102.
  
 
\bibitem{Sonoda2007} H. Sonoda, G. Watanabe, K. Sato, K. Yasuoka, and T. Ebisuzaki, Phys. Rev. C {\bf 77}, (2008) 035806, [Erratum: Phys.Rev.C {\bf 81}, (2010) 049902].
\bibitem{Avancini2010} S. S. Avancini, S. Chiacchiera, D. P. Menezes, and C. Provid\^encia, Phys. Rev. C {\bf 82}, (2010) 055807, [Erratum: Phys.Rev.C {\bf 85}, (2012) 059904].
\bibitem{Avancini2017} S. S. Avancini, M. Ferreira, H. Pais, C. Provid\^encia, and G. R\"opke, Phys. Rev. C {\bf 95}, (2017) 045804.
\bibitem{Ji2020} F. Ji, J. Hu, S. Bao, and H. Shen, Phys. Rev. C {\bf 102}, (2020) 015806.

\bibitem{Wu2017} Xin-Hui Wu, Si-Bo Wang, Armen Sedrakian, and Gerd R\"opke, J. Low Temp. Phys. {\bf 189}, (2017) 133.
\bibitem{Sedrakian2020} A. Sedrakian, Eur. Phys. J. A {\bf 56}, (2020) 258.

\bibitem{Fischer20} T. Fischer \textit{et al.}, arXiv:2008.13608 [astro-ph.HE].
\bibitem{rosswog15}  S. Rosswog,  Int. J. Mod. Phys. D {\bf 24}, (2015) 1530012.
\bibitem{avancini10} S. S. Avancini, C. C. Barros Jr., D. P. Menezes, and C. Provid\^encia,
Phys. Rev. C {\bf 82}, (2010) 025808.
\bibitem{avancini12} S. S. Avancini, C. C. Barros, Jr., L. Brito, S. Chiacchiera, D. P. Menezes, and C. Provid\^encia, Phys. Rev. C {\bf 85}, (2012) 035806.
\bibitem{typel10} S. Typel, G. R\"opke, T. Kl\"ahn, D. Blaschke and H. H. Wolter, Phys. Rev. C {\bf 81}, (2010) 015803.
\bibitem{ferreira12} M. Ferreira and C. Provid\^encia, Phys. Rev. C {\bf 85}, (2012) 055811.  
\bibitem{pais15} H. Pais, S. Chiacchiera, and C. Provid\^encia, Phys. Rev. C {\bf 91}, (2015) 055801.
\bibitem{avancini17} S. S. Avancini, M. Ferreira, H. Pais, C. Provid\^encia, and G. R\"opke, Phys. Rev. C {\bf 95}, (2017) 045804. 
\bibitem{PaisPRC97} H. Pais, F. Gulminelli, C. Provid{\^e}ncia, and G. R{\"o}pke, Phys. Rev. C {\bf 97} (2018) 045805. 
\bibitem{PaisPRC2019} H. Pais, F. Gulminelli, C. Provid{\^e}ncia, and G. R{\"o}pke, Phys. Rev. C {\bf 99}, (2019) 055806.
\bibitem{FSU} B. G. Todd-Rutel and J. Piekarewicz, Phys. Rev. Lett. {\bf 95}, (2005) 122501.
\bibitem{tm1} Y. Sugahara, and H. Toki, Nucl. Phys. A {\bf 579}, (1994) 557.
\bibitem{shen} H. Shen, H. Toki, K. Oyamatsu, and K. Sumiyoshi, Nucl. Phys. A {\bf 637}, (1998) 435.
\bibitem{providencia13} C. Provid\^encia and A. Rabhi,
Phys. Rev. C {\bf 87}, (2013) 055801.
\bibitem{bao14} S. S. Bao, and H. Shen, Phys. Rev. C {\bf 89}, (2014) 045807. 
\bibitem{pais16} H. Pais and C. Provid\^encia, Phys. Rev. C {\bf 94}, (2016) 015808.
\bibitem{shen2020} Hong Shen, Fan Ji, Jinniu Hu, and Kohsuke Sumiyoshi,  Astrophys. J. {\bf 891}, (2020) 148. 
\bibitem{nl3} G. A. Lalazissis, J. Konig, and P. Ring, Phys. Rev. C {\bf 55}, (1997) 540.
\bibitem{horowitz01} C. J.  Horowitz and J. Piekarewicz, Phys. Rev. Lett. {\bf 86}, (2001) 5647.
\bibitem{sfho} Andrew W. Steiner, Matthias Hempel, and Tobias Fischer, Astrophys.  J. {\bf 774}, (2013) 17. 
\bibitem{FSU2R} L. Tolos, M. Centelles, and A. Ramos, Pub. Astron. Soc. Aust. {\bf 34}, (2017) e065.
\bibitem{ddme2} G. A. Lalazissis, T. Nik\v si\'c, D. Vretenar, and P. Ring, Phys. Rev. C {\bf 71}, (2005) 024312.
\bibitem{tsang12} M. Tsang \textit{et al.}, Phys. Rev. C {\bf 86}, (2012) 015803.
\bibitem{lim13} J. M. Lattimer and Y. Lim, Astrophys. J. {\bf 771}, (2013) 51.
\bibitem{oertel18} M. Oertel, M. Hempel, T. Kl\"ahn, and S. Typel, Rev. Mod. Phys. {\bf 89}, (2017) 015007.
\bibitem{hebeler13} K. Hebeler,  J. Lattimer, C. Pethick, and A. Schwenk, Astrophys. J. {\bf 773}, (2013) 11.
\bibitem{qin12} L. Qin, K. Hagel, R. Wada, J. B. Natowitz, S. Shlomo, A. Bonasera, G. R\"opke, S. Typel, Z. Chen, M. Huang, \textit{et al.}, Phys. Rev. Lett. {\bf 108}, (2012) 172701.
\bibitem{indra} R. Bougault \textit{et al.}, J. Phys. G {\bf 47}, (2020) 025103.
\bibitem{PaisPRL} H. Pais, R. Bougault, F. Gulminelli, C. Provid\^encia, \textit{et al.}, Phys. Rev. Lett. {\bf 125}, (2020) 012701.
\bibitem{PaisJPG} H. Pais, R. Bougault, F. Gulminelli, C. Provid\^encia, \textit{et al.}, J. Phys. G: Nucl. Part. Phys. {\bf 47}, (2020) 105204.
\bibitem{Natowitz20} J. B. Natowitz, H. Pais, G. R\"opke, \textit{et al.}, arXiv:2009.05200 [nucl-ex].
\bibitem{Aymard14} F. Aymard, F. Gulminelli, and J. Margueron, Phys. Rev. C {\bf 89}, (2014) 065807.
\bibitem{raduta2} F. Gulminelli and Ad. R. Raduta, Phys. Rev. C {\bf 92}, (2015) 055803.
\bibitem{Roepke15} G. R\"opke, Phys. Rev. C {\bf 92}, (2015) 054001.
\bibitem{virial1} C. Horowitz and A. Schwenk, Nucl. Phys. A {\bf 776}, (2006) 55.
\bibitem{virial2} M. Voskresenskaya and S. Typel, Nucl. Phys. A {\bf 887}, (2012) 42.
\bibitem{yudin19} A. V. Yudin, M. Hempel, S. I. Blinnikov, D. K. Nadyozhin, and I. V. Panov, Mon. Not. Roy. Astron. Soc. {\bf 483}, (2019) 5426.
\bibitem{Roepke20} G. R\"opke, Phys. Rev. C {\bf 101}, (2020) 064310.

\bibitem{olfa20} Olfa Boukari, Helena Pais, Sofija Anti\'c, and Constan\c ca Provid\^encia, arXiv:2007.08852 [nucl-th].
\bibitem{hempel2015} M. Hempel, K. Hagel, J. Natowitz, G. R\"opke, and S. Typel,
Phys. Rev. C {\bf 91}, (2015) 045805.


\end{thebibliography}
\end{document}